\documentclass[]{aa}
\usepackage{graphicx}
\usepackage[intlimits,sumlimits]{amsmath}
\usepackage{times,epsfig} 
\usepackage{natbib}
\usepackage{amssymb}
\bibpunct{(}{)}{;}{a}{}{,} 
\usepackage{amsmath}
\usepackage[english]{babel}

\newcommand{\kms}{${\rm km~s^{-1}~Mpc^{-1}}$}

\begin{document}
\title{Prospects and Pitfalls of Gravitational Lensing in Large
  Supernova Surveys}
\titlerunning{Prospects and Pitfalls of Gravitational Lensing in Large
SN Surveys}
\authorrunning{J\"onsson}
\author{
J.~J\"onsson\inst{1,2}
 	\and T. Kronborg\inst{3,4}
 	\and E. M\"ortsell\inst{1}
	\and J. Sollerman\inst{1,3}
}

\institute{Stockholm Observatory, AlbaNova, Department of Astronomy, S-10691 
Stockholm, Sweden 
\and
 University of Oxford Astrophysics, Denys Wilkinson Building, Keble Road, Oxford OX1 3RH, UK 
\and
Dark Cosmology Centre, Niels Bohr
Institute, University of Copenhagen, Juliane Maries Vej 30, DK-2100 
Copenhagen \O, Denmark
\and
LPNHE, CNRS-IN2P3 and Universit\'es Paris VI \& VII, 4 place Jussieu, 75252 Paris Cedex 05, France \\
\email{jacke@astro.ox.ac.uk}
} 
\offprints{J. J\"onsson}
\date{Received -- / Accepted --}
\abstract{}
{To investigate the effect of gravitational lensing of supernovae in large
  ongoing surveys.}
{We simulate the effect of gravitational lensing magnification on
  individual supernovae using observational data input from two large
  supernova surveys. To estimate
  the magnification due to matter in the foreground, we simulate galaxy 
  catalogs and compute the magnification along individual lines of
  sight using the multiple lens plane algorithm. The dark matter haloes
  of the galaxies are modelled as gravitational lenses 
  using singular isothermal sphere or Navarro-Frenk-White profiles. 
  Scaling laws between luminosity and mass,
  provided by Faber-Jackson and Tully-Fisher relations, are used to
  estimate the masses of the haloes.  
}
{
  While our simulations show that the SDSSII supernova survey is
  marginally affected by gravitational lensing, we find that the
  effect will be measurable in the SNLS survey that probes higher
  redshifts.
  Our simulations show that the probability to measure a
  significant ($3\sigma$) correlation between the Hubble diagram residuals 
  and the calculated lensing magnification is $\ga 95\%$ in the SNLS data.
  Moreover, with this data it should be possible to constrain the 
  normalisation of the 
  masses of the lensing galaxy haloes at the
  $1\sigma$ and $2\sigma$ confidence level with $\sim 30\%$ and $\sim 60\%$ 
  accuracy, respectively.
}
{}

\maketitle
\section{Introduction}
Type Ia supernovae (hereafter SNe~Ia) are exceptionally useful tools
for cosmological investigations. Ongoing large supernova surveys, such
as ESSENCE \citep{miknaitis07,wv07}, SNLS \citep{astier06} and SDSSII
\citep{frieman07} harvest hundreds of distant supernovae with high
precision.

Several effects will alter the luminosity of these distant sources,
and thus dilute the cosmological signal. In this paper we investigate
gravitational lensing (de)magnification of the light from these
explosions. Although this effect is hardly large enough to dominate
the uncertainties in current experiments, as has already been
addressed by several groups on statistical grounds
\citep[e.g.,][]{riess04,hl05,astier06,wv07}, the question if it can be
corrected for on an individual supernova basis remains. This has been
investigated in a series of papers \citep{gunnarsson06,jonsson06} and
a tentative detection of a correlation between calculated lensing
magnification and Hubble residuals was recently found using the very
high-$z$ supernovae in the GOODS field \citep{jonsson07}.  

In this paper we investigate to what extent other surveys can be
affected. 
The GOODS supernovae \citep{riess04,strolger04,riess07}
remain unique in that their large distances clearly make them more
likely to be lensed. On the other hand, the ongoing ground based
surveys will measure significantly more supernovae with much better
sampling, and the improved statistics may well compensate for the
smaller distances. To find out which surveys are most likely to
display a lensing signal we have performed simulations based on our
previous work. The results are presented below, and show that the
prospects of detecting lensing are very good. 

In Sect.~\ref{surveys}
we briefly introduce the surveys we have investigated and the data
from these surveys that are used as input for the simulations. A short
summary of gravitational lensing of SNe~Ia is given in
Sect.~\ref{lensing}. Section~\ref{simulations} describes the
simulations and the results are presented in Sect.~\ref{results},
where we also discuss how a detection of a lensing signal could be
used to
obtain information about the
lensing matter. Finally, we summarise our results in
Sect.~\ref{discussion}.

\section{Supernova surveys}\label{surveys}
There are (at least) three ongoing major ground based supernova
surveys as mentioned above. We have chosen to simulate the SNLS survey
to estimate the effects in a deep survey. It probes a similar redshift
range as the ESSENCE survey and is likely to find more supernovae. It
is also of importance that SNLS uses several filters which allows
photo-$z$ determinations for the field galaxies.
The SDSSII supernova survey is selected to investigate if any
gravitational lensing effects could be expected for this smaller
redshift domain when we have good statistics.

\subsection{SNLS}
The Supernova Legacy Survey, \citep[SNLS,][]{astier06} has published
constraints on the cosmological parameters using 71 SNe~Ia 
ranging from $z=0.2-1.0$ discovered during the first year. Around 500
spectroscopically identified SNe Ia are expected to end up in the
final Hubble diagram at the end of the survey.  The survey consists of
an imaging survey detecting and monitoring the light curves of the
SNe, and a spectroscopic follow-up confirming the nature of the
SNe and measuring their redshifts. A wide field imager used in
"rolling search" mode allows to perform the discovery and photometric
follow-up at the same time. The SNLS is searching for SNe in 4
different fields of 1 square degree each and each field is observed
every third to fourth night for as long as it remains
visible. Observations are taken in 4 filters (the MegaCam filter set)
giving rise to a promising data set for observing the lensing signal.

Since the line of sight to SNe~Ia located close to the edge of the
fields can not be modelled for gravitational lensing accurately, we
estimate that only $\sim 450$ of the $\sim 500$ Hubble diagram SNe~Ia
can be used to search for a gravitational lensing signal. The
parameters from this survey that are needed as input for this
investigation
\citep[see][]{gunnarsson06} have been taken from the 
literature \citep{ilbert06} and from discussions with the involved 
astronomers. 

\subsection{SDSSII}
The Sloan Digital Sky Survey II Supernova Survey
\citep[SDSSIISN,][]{frieman07} is searching a 280 square degree field
over 3 seasons. The survey is well on track to discover over 500 Type
Ia supernovae, of which $\sim300$ will likely end up in the Hubble
diagram. The SDSSII is a follow-up on the previous successfull SDSS
projects, which means that this search has a working infrastructure
capable of handling large amounts of data. The search is performed in
5 bands ($ugriz$) and the SDSS field is calibrated down to an accuracy
of $\sim1\%$ \citep{ivezic07}.

\section{Gravitational lensing of SNe~Ia}\label{lensing}
Here we give a short introduction to the aspects of gravitational
lensing of SNe~Ia which are relevant to this work. For a more thorough
introduction to the subject see e.g.~\cite{bergstrom00}. 

Matter inhomogeneities in the Universe give rise to gravitational
fields which will influence the path of light rays propagating between
a source and an observer. In the weak lensing regime, gravitational
bending of light, or gravitational lensing, can lead to changes in the
apparent position of the source as well as changes in the flux.
For point sources such as SNe~Ia, the amplification (or
de-amplification) of the flux can be described by the magnification
factor, $\mu$. The observed flux from a point source in an
inhomogeneous universe is given by $f_{\rm obs}=\mu f$, where $f$ is
the flux that would be observed in a homogeneous universe in the
absence of lensing.  Amplification and de-amplification relative to a
homogeneous universe is consequently described by $\mu>1$ and $\mu<1$,
respectively.

The distribution of magnification factors, $P(\mu)$, for sources at a
specific redshift depends on the distribution of matter between the
source and the observer. In general $P(\mu)$ is asymmetric with a peak
at $\mu<1$ and a high magnification tail. Due to flux conservation, as
long as we do not have multiple images, the average magnification is
unity, i.e.~$\langle\mu\rangle=\int P(\mu)
\mu d\mu=1$. The average flux from a number of standard candles, all
at the same redshift, $\langle f_{\rm obs}\rangle= \langle \mu\rangle
f=f$, is thus an unbiased distance indicator if the variance of the
flux is finite. However, changes in the flux from standard candles
will increase the dispersion in distance measurements and could
potentially lead to selection effects.

\section{Simulations}\label{simulations}
We use simulations to investigate the effects of gravitational lensing
for the surveys described in Sect.~\ref{surveys}.  Since gravitational
lensing depends on the distribution of matter between the source and
the observer, a model of the matter distribution is needed in order to
simulate gravitational magnification. In the next section we briefly
describe our model of the matter distribution in the Universe, which
is based on observational input. For more details we refer the reader
to \citet{gunnarsson06} where our model is described in detail.

\subsection{Simulating lines of sight}\label{simlos}
For each simulated SN~Ia we simulate a line of sight populated by
foreground galaxies. The galaxy population is characterised by the
luminosity functions presented in \cite{dahlen05}. These luminosity
functions are used to simulate the number of galaxies along a line of
sight as well as their brightness and type. Simulated $B$-band
absolute magnitudes of the galaxies are in the range $-23<M_B<-16$.  A
constant comoving number density of galaxies with redshift is
assumed. Since the position of the galaxies in the sky as well as the
redshifts are assigned at random, clustering of matter into larger
structures is not taken into account. 

Each galaxy is associated with a dark matter halo and we assume that
the mass of the halo can be obtained from the luminosity of the
galaxy. Gravitational lensing depends not only on the mass of the lens,
but also on the distribution of matter within the lens. 
We will model the galaxy haloes using 
singular isothermal sphere (SIS) and 
Navarro-Frenk-White \citep[NFW,][]{navarro97} profiles. Since these
profiles are divergent, they are truncated at $r_{200}$, the radius
inside which the mean density is 200 times the critical density of the 
Universe. We use numerical routines \citep{navarro97} to find the
concentration of a halo characterised by the mass
$m_{200}$ enclosed within the radius $r_{200}$. 

To convert galaxy luminosity to halo mass we relate dynamical
properties of the galaxies to the dark matter halo. For a SIS halo,
$m_{200}$ is related to the velocity dispersion, $\sigma$, via the
formula
\begin{equation}
m_{200}=\frac{\sqrt{2}\sigma^3}{5GH_0}.
\label{eq:m200}
\end{equation}
This formula is a good approximation also for a NFW halo with the same 
corresponding maximum rotation velocity, $V_{\rm max}$. 
The mass of a NFW halo is
over-estimated by $5-15\%$ when Eq.~(\ref{eq:m200}) is used to
calculate $m_{200}$ for a given value of $V_{\rm max}$.
The velocity dispersion of early type galaxies
can be obtained from the absolute magnitudes through the
Faber-Jackson (F-J) relation as derived by \cite{mitchell05}
\begin{equation}
\log \sigma=-0.091(M_B-4.74+0.85z).
\label{eq:FJ}
\end{equation}
The redshift dependence in the relation accounts for the brightening
of the stellar population with redshift. We represent the measurement
error in this relation by the scatter in the SDSS measurements reported
by \cite{sheth03}
\begin{equation}
{\rm rms}(\log \sigma)=0.079[1+0.17(M_B+19.705+0.85z)].
\label{eq:FJrms}
\end{equation}
For late type galaxies we use the Tully-Fisher (T-F) relation 
derived by \cite{pierce92}, 
with correction for redshift calculated by \cite{bohm04},
\begin{equation}
\log V_{\rm max}=-0.134(M_B+3.61+1.22z),
\label{eq:TF}
\end{equation}
to relate galaxy luminosity to the maximum
rotation velocity of the galaxy. The velocity
dispersion relates to the maximum rotation
velocity via $\sigma=V_{\rm max}/\sqrt{2}$.
For the T-F relation the observed scatter is 
${\rm rms}(M_B)=0.41$, which corresponds to
\begin{equation}
{\rm rms}(\log V_{\rm max})=0.06.
\label{eq:TFrms}
\end{equation}
The mass range for early and late type galaxies computed using the
formulae above is 
$4\times10^{11}<m_{200}/M_{\sun}<3\times10^{13}$ and
$3\times10^{10}<m_{200}/M_{\sun}<2\times10^{13}$, respectively.

The code used to compute the magnification factor for the simulated
lines of sight is described in the following subsection.

\subsection{Q-LET}
The Quick Lensing Estimation Tool \citep[Q-LET,][]{gunnarsson04} is
a numerical code which utilises the multiple lens plane algorithm. 
Each dark matter halo is projected into a lens plane perpendicular to the
line of sight situated at the angular diameter distance corresponding
to the redshift of the galaxy. 
The deflection angle is then computed for each lens plane and a ray
originating at the position of the image is traced back to the source
position via the lens equation. From the Jacobian determinant of the
lens equation, the magnification factor can be computed.  

In our model universe, the matter density of dark matter galaxy
haloes, $\rho_{\rm g}(z)$, is typically less than the measured global
matter density $\rho_{\rm m}(z)=\Omega_{\rm M}\rho_{\rm c}^0(1+z)^3$,
where $\rho_{\rm c}^0$ is the present critical density of the
Universe.  Since only galaxies brighter than the magnitude limit will be
associated with haloes, we expect $\rho_{\rm g}(z)$ to decrease with
redshift.  To ensure that our model is consistent, we describe the
matter which is ``missing'' due to our ignorance as a smoothly
distributed component characterised by the smoothness parameter given
by $\eta(z)=1-\rho_{\rm g}(z)/\rho_{\rm m}(z)$\footnote{A completely
smooth universe and its opposite, a completely clumpy universe,
corresponds to $\eta=1$ and $\eta=0$, respectively.}.

Numerical routines in \citet{kayser97}, which can handle a redshift
dependent smoothness parameter, are used to calculate the 
 angular diameter distances in a clumpy universe which are involved in 
the gravitational 
lensing calculations\footnote{The cosmological model used in the 
calculations is described
by $\Omega_{\rm M}=0.3$, $\Omega_{\Lambda}=0.7$, and $H_0=70$ \kms.}.

\subsection{Estimating magnification factors}\label{uncertainties}
It should be possible to detect gravitational lensing of SNe~Ia through
the expected correlation between $f_{\rm obs}$ and $\mu$. However,
since $\mu$ is not observable we have to use an estimate, $\mu_{\rm
  est}$, when we search for the correlation. 
To simulate the precision to which
magnification factors can be estimated, we first compute $\mu$ for a fiducial
model and then try to recover the value of the magnification factor 
in the presence of different uncertainties.
These uncertainties, which are described in the following paragraphs, 
dilutes the lensing signal carried by the estimated magnification
factor $\mu_{\rm est}$.

\paragraph{Finite field size}
Only a limited number of
foreground galaxies can be taken into
account in the calculations, since the fields are of finite size. 
To limit this source of error to the percent level, all galaxies within
$60\arcsec$ should be included \citep{gunnarsson06}. To include this
source of error in our simulations a larger number of galaxies is
included in the calculation of $\mu$ (all galaxies within
$100\arcsec$) than in the calculation of
$\mu_{\rm est}$ (all galaxies within $60\arcsec$).

\paragraph{Photometric redshifts} 
Only galaxies brighter than the magnitude limit can be included in our
calculations of the estimated magnification factors, $\mu_{\rm
est}$. For surveys like SNLS we have to rely upon photometric
redshifts in our calculations. The magnitude limit of the SNLS is 
$i'_{\rm AB}=25.5$, but
reliable photometric redshifts can only be obtained to $i'_{\rm
AB}=24$, which we take as the magnitude limit in our simulations. 
Photometric redshifts, $z_{\rm p}$, usually have larger
uncertainties than spectroscopic redshifts, $z_{\rm s}$. More accurate
photometric redshifts can be obtained for bright galaxies than for
faint galaxies. According to \cite{ilbert06} the photometric redshift
accuracy of galaxies observed within the CFHT legacy survey with
brightness $17.5 < i'_{\rm AB} < 22.5$ and $22.5 < i'_{\rm AB} < 24$
is $\sigma_{\Delta z/(1+z)}=0.029$ and $\sigma_{\Delta
z/(1+z)}=0.034$, respectively.

For some fraction, $f$, of the galaxies, the calculation of the
photometric redshift fails miserably. These erroneous - or
``catastrophic'' - redshifts often arise due to confusion between the
Lyman break ($\lambda_{\rm L}=1215$~\AA) and the Balmer break
($\lambda_{\rm B}=4000$~\AA). We use a simple model of catastrophic
redshifts based on this confusion and the results presented in
\cite{ilbert06}.  When the Lyman and Balmer breaks are confused either
the condition
\begin{equation}  
(1+z_{\rm p})\lambda_{\rm B}=(1+z_{\rm s})\lambda_{\rm L},
\label{eq:phot1}
\end{equation}
or the condition 
\begin{equation}  
(1+z_{\rm p})\lambda_{\rm L}=(1+z_{\rm s})\lambda_{\rm B},
\label{eq:phot2}
\end{equation}
is fulfilled. If the condition expressed by Eq.~(\ref{eq:phot1})
is fulfilled, the
photometric redshift is under-estimated. This condition can only be
fulfilled if $z_{\rm s}>2.3$. On the other hand, the photometric 
redshift is over-estimated if Eq.~(\ref{eq:phot2}) is fulfilled. 
According to Fig.~6c in \cite{ilbert06} over-estimation of $z_{\rm p}$
is $\sim 5$ times more frequent than under-estimation. 
In our simple model a fraction $f=1.9\%$ \citep{ilbert06} 
of the bright galaxies 
($17.5 < i'_{\rm AB} < 22.5$) are assigned catastrophic redshifts.  
The proportion of catastrophic redshifts computed using Eq.~(\ref{eq:phot1})
and Eq.~(\ref{eq:phot2}) is 1:5.
For the faint galaxies 
($22.5 < i'_{\rm AB} < 24$) 
we use the same model, but
the fraction of catastrophic redshifts is 
higher, $f=5.5\%$ \citep{ilbert06}.

\paragraph{Scatter in the F-J and T-F relations}
In the conversion between galaxy luminosity and halo mass, the
Faber-Jackson and Tully-Fisher relations are used. The scatter in
these relations,
given by Eq.~(\ref{eq:FJrms}) and Eq.~(\ref{eq:TFrms}), 
are the largest sources of uncertainty in the calculations.
When $\mu_{\rm est}$ is
estimated this scatter is taken into account. 
We also investigate the effects of systematic errors in the
relations between luminosity and mass (see Sect.~\ref{haloconstraints}). 
Missclassification of galaxy type is a related potential uncertainty
which we have not investigated. However, since the missclassification
rate for the brightest galaxies,
which are also the most important lenses, is likely \citep{dah08} to be
only a few percent, we assume the neglection of this source of
uncertainty to be safe. 

\paragraph{Halo models}
For the fiducial model, all galaxies are described by NFW profiles.
When we estimate the uncertainties in the magnification factor we 
assume incorrect halo
models, i.e.~SIS instead of NFW profiles, for 50\% of the lenses.

\subsection{Correlation}\label{sec:correlation}
To predict the probability of detecting a correlation 
between estimated magnifications and Hubble diagram residuals
we use Monte
Carlo simulations. A large number of SNe~Ia data sets are simulated and
for each data set the linear correlation coefficient, $r$, is computed. 
The residuals in the Hubble diagram, 
which is built of SNe~Ia corrected for light 
curve shape and color, depends on the \emph{true} 
magnification factor (computed using the fiducial model),
\begin{equation}
\Delta=-2.5 \log_{10}\mu.
\label{eq:residual}
\end{equation}

In our simulations we add gaussian
noise to the Hubble diagram residuals from 
intrinsic brightness scatter and measurement errors.
For the intrinsic dispersion we use the value derived by
\cite{astier06}, $\sigma_{\rm int}=0.13$ mag. 
To model the measurement errors in the SNLS we use the following redshift
dependent fit to the measurement errors in \citet{astier06} 
\begin{equation}
\sigma_{\rm err}=\left\{ 
\begin{array}{ll}
 0.05 \mbox{ mag} &\mbox{ if $z<0.8$} \\
 0.84z^2-1.04z+0.34 \mbox{ mag}  &\mbox{ if $z \geq 0.8$}.
       \end{array}
\right.
\end{equation}
The total noise in the
residuals are given by $\sigma_{\rm int}$ and $\sigma_{\rm err}$ added
in quadrature.
This noise or other sources of noise could in reality be non-gaussian. 
Such non-gaussianity will be a concern for 
techniques that try to detect gravitational lensing of SNe~Ia based on the 
asymmetry of the distribution of residuals \citep[e.g.,][]{wan05}. However, for this study where we calculate the magnification for each individual supernova, 
as long as the contributions to the residuals are not correlated with the foreground matter,
the gravitational lensing signal should not be systematically distorted, only diluted.

The magnification, in logarithmic units, on the other hand,
depends on the \emph{estimated} magnification factor,
\begin{equation}
\Delta_{\mu}=-2.5 \log_{10}\mu_{\rm est}.
\label{eq:magnification}
\end{equation}
The uncertainties contributing to the noise in the
estimated magnification factors were described in Sect.~\ref{uncertainties}.

Since we assume the intrinsic brightness scatter and measurement errors
to be gaussian and these errors dominate, the distributions of 
correlation coefficients are gaussian as well and can be described by the
mean value, $\langle r \rangle$, and the standard deviation, $\sigma_r$. 
To evaluate the confidence level of the detection of the correlation,
we compare with what we could expect to measure if there is no
correlation. For each simulated data set we compute the correlation
coefficient of the null hypothesis, $r_{\rm null}$, i.e.~the
hypothesis of no lensing.  Since both distributions are gaussian, the
probability of a $n\sigma$ detection is given by
\begin{equation}
P(n\sigma)=\frac{1}{2}{\rm erfc}\left( \frac{\langle r_{\rm null} \rangle+
n\sigma_{r_{\rm null}}-\langle r \rangle}{\sqrt{2}\sigma_r} \right),
\end{equation}
where $\langle r_{\rm null} \rangle$ and $\sigma_{r_{\rm null}}$ 
refers to the mean and
standard deviation of the distribution of $r_{\rm null}$, respectively.

\section{Results}\label{results}

\subsection{Gravitational magnification}
In this section we present some basic properties of the simulated gravitational
magnification distributions.

\subsubsection{SNLS}
We have simulated the magnification distribution for sources
following the redshift distribution of the first year 
SNLS SNe~Ia \citep{astier06}.
The magnification factors used to build up this composite distribution 
were drawn from redshift dependent
magnification distributions, which were all normalised to fulfil the
condition $\langle \mu \rangle=1$.
Although the average magnification factor is unity,
the average value of $\Delta$, which depends on the higher moments of the
magnification distribution, 
deviates 
from zero, $\langle \Delta \rangle=0.0011$ mag. 
Of importance is also the median of the magnification distribution, 
$\Delta_{1/2}=0.0084$ mag,
which tells us that 
most SNe~Ia in the SNLS data set are slightly
demagnified, and would consequently appear to be more distant than they
really are.

\subsubsection{SDSSII}\label{sec:SDSSIIa}
According to our simulations, 
the average and the median of the composite magnification distribution 
corresponding to the SDSSII SN~Ia survey is
$\langle \Delta \rangle =0.00018$ mag and  
$\Delta_{1/2}=0.00080$ mag, respectively. The magnification of most
SNe~Ia are thus clearly negligible for the SDSSII, which is of course a
consequence of the shallower redshifts probed by this survey. 
Since the expected effect of gravitational lensing is very small for
the SDSSIISN, we focus on gravitational lensing of SNLS SNe~Ia for the
rest of this paper.

\subsection{Shifts in cosmological parameters}
Gravitational lensing can potentially lead to biased cosmological
results. 
Here we investigate and quantify these effects for the
SNLS and show that the effects are indeed rather small.
We have studied shifts in cosmological parameters for two cases:
\begin{enumerate}
\item The universe is assumed to be flat and dominated by a
  cosmological constant. In this case the present matter density, 
  $\Omega_{\rm M}$ is fitted to simulated data sets.
\item The universe is assumed to be flat and the dark energy equation
  of state parameter, $w$, is furthermore assumed to be constant. 
  In this case $\Omega_{\rm M}$ and $w$ are simultaneously fitted to data and 
  constrains from Baryonic Acoustic Oscillations
 \citep[BAO,][]{eisenstein05} are used.
\end{enumerate}
A simulated low redshift data set consisting of 44 SNe~Ia, 
with properties similar to that
used by \citet{astier06}, is used to anchor the Hubble diagram.

We characterise the shift in $\Omega_{\rm M}$ and $w$ by the
differences, 
$\Delta \Omega_{\rm M}=\Omega_{\rm M}^{\rm lensed}-\Omega_{\rm M}$ and 
$\Delta w=w^{\rm lensed}-w$, in the best fit values obtained with and 
without gravitational lensing. Our results are collected in 
Table~\ref{tab:bias}.
\begin{table*}
\begin{minipage}[t]{\columnwidth}
\caption{Average and root mean square of distributions 
of $\Delta \Omega_{\rm M}$  
and $\Delta w$ 
for case 1 and case 2 for different SNLS like data sets
  with and without bright outliers removed.}         
\label{tab:bias}      
\centering                                      
\renewcommand{\footnoterule}{}  
\begin{tabular}{ccllll}          
\hline\hline                        
& Outliers\footnote{Bright SNe~Ia deviating more than
  $2.5\sigma$ from the best fit cosmology.} & \multicolumn{2}{c}{Case 1} & 
\multicolumn{2}{c}{Case 2} \\
$N$  & removed& $\langle \Delta \Omega_{\rm M} \rangle $ & $ 
\sigma_{\Delta \Omega_{\rm M}}$ &
$\langle \Delta w \rangle $ & $
\sigma_{\Delta w}$ \\
\hline
 70 & no &$-0.0013 $ & $0.0075$ &
$-0.0020 $ & $ 0.013$ \\ 
 70 & yes & $-0.0038 $ & $ 0.010$ &
$-0.0043 $ & $ 0.013$ \\
 500 & no & $-0.0016 $ & $0.0045$ &
$-0.0023 $ & $ 0.0059$ \\
 500 & yes & $-0.0052 $ & $ 0.0061$ & 
$-0.0051 $ & $ 0.0056$ \\
\hline                                             
\end{tabular}
\end{minipage}
\end{table*}

Since flux is conserved, effects of gravitational lensing will average
out for large numbers of SNe~Ia sampling the whole magnification
distribution. In this study, however, magnitudes are used instead of
fluxes, as is common in supernova cosmology. When magnitudes are used,
the average magnitude is shifted compared to the case of no lensing.
This shift is caused by higher order moments of the magnification 
factor distribution, which are responsible for the fact that
$\langle \Delta \rangle \neq 0$ even though $\langle \mu \rangle=1$.
We thus expect a small shift in the cosmological
parameters solely due to our choice of using magnitudes. 
For the SNLS, where $\langle \Delta \rangle=0.0011$ mag, these shifts are
expected to be 
$\Delta \Omega_{\rm M}=-0.0012$ for case 1 and 
$\Delta \Omega_{\rm M}=-0.00030$ and 
$\Delta w=-0.0024$ for case 2.

\begin{figure}
\resizebox{\hsize}{!}{\includegraphics[angle=-90]{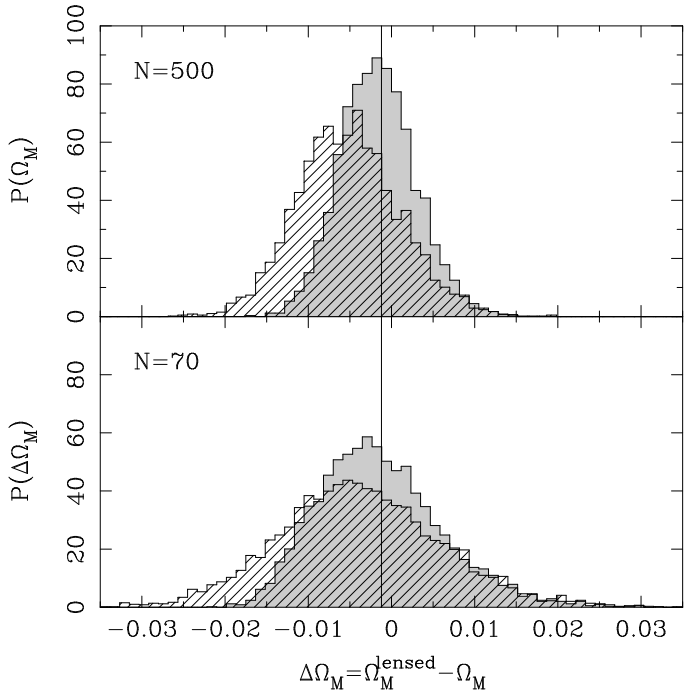}}
\caption{\label{fig:ombias} Shift in $\Omega_{\rm M}$ due to
  gravitational lensing. The grey shaded distributions show 
$\Delta \Omega_{\rm M}=\Omega_{\rm M}^{\rm lensed}-\Omega_{\rm M}$ 
for $\sim 5000$ simulated data sets consisting of 70 
(bottom panel) and 500 (top panel) SNLS SNe~Ia. The vertical solid line
shows the shift ($\Delta \Omega_{\rm M}=-0.0012$) expected due to 
higher order moments of the magnification distribution. 
The hatched histograms show the shift which would be the result of
removing bright $2.5\sigma$
outliers from the Hubble diagram. 
}
\end{figure}

Figure~\ref{fig:ombias} shows the shift in $\Omega_{\rm M}$ 
which, according to or simulations, can be expected for 
SNLS first year and final data sets. The
top and bottom panel shows the distribution (shaded histograms)
 of $\Delta \Omega_{\rm M}$ for 70 and 500 SNe~Ia, respectively. 
Here we consider the final data set which will be used for
cosmology and not the smaller subset which could be used to detect lensing.
For $N=70$ the distribution is clearly asymmetric, while for $N=500$
the distribution is fairly gaussian. 
The average shift, $\langle \Delta \Omega_{\rm M} \rangle$, or bias,
is very close to the value expected for both $N=70$ and $N=500$. 

\cite{astier06} estimate $\Omega_{\rm M}$ obtained from the first year
SNLS data set to be shifted due to gravitational lensing 
by at most $\Delta \Omega_{\rm M}=-0.005$. 
The distribution in Fig.~\ref{fig:ombias} peaks close to this value 
for $N=70$. 
We expect the contribution to the uncertainty in $\Omega_{\rm M}$ 
from gravitational lensing 
to be of the order $0.008$ or
$3\%$ at the $1\sigma$  ($68.3\%$) confidence level. Much of the 
dispersion due to gravitational lensing should, however,
already be included in the SNLS analysis 
via the intrinsic dispersion derived from the data themselves. 
For the final data set, which will consist of $\sim 500$ SNe~Ia, 
the expected contribution to the error in $\Omega_{\rm M}$ from lensing 
is $\la 2\%$ at the $1\sigma$ confidence level.

\begin{figure}
\resizebox{\hsize}{!}{\includegraphics[angle=-90]{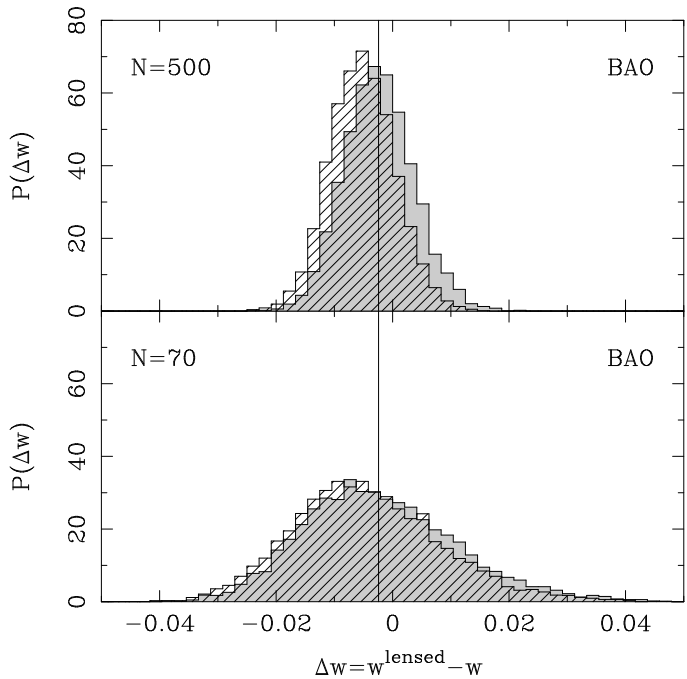}}
\caption{\label{fig:bias} Shift in $w$ due to
  gravitational lensing. 
  The grey shaded distributions show
  $\Delta w=w-w^{\rm lensed}$
  for $\sim 5000$ simulated data sets consisting of 70 
  (bottom panel) and 500 (top panel) SNLS SNe~Ia. The vertical solid line
  shows the shift ($\Delta w=-0.0024$) expected due to 
  higher order moments of the magnification distribution. 
  The hatched histograms show the shift which would be the result of
  removing bright $2.5\sigma$
  outliers from the Hubble diagram. Constraints from BAO were used in the
  cosmology fits.
}
\end{figure}

Let us now consider the shift for case 2. Figure~\ref{fig:bias} shows the
results of our simulations for SNLS. 
To each simulated data set, $\Omega_{\rm M}$ and $w$ were fitted 
simultaneously with and without
lensing. The shaded histograms in the top and bottom panel in 
Fig.~\ref{fig:bias} show the probability distributions of $\Delta w$
for $N=500$ and $N=70$, respectively. Since the shift in 
$\Omega_{\rm M}$ is very small for case 2, we only consider $\Delta w$.
Also for case 2 the expected bias, $\langle \Delta w \rangle$, is
close to what we expect due to the
higher order moments in the magnification factor distribution 
(outlined by the vertical lines in the figure).

The most likely shift for $N=70$ is $\Delta w \approx -0.005$, 
which is half the maximum shift
($\Delta w=-0.01$) estimated by \citet{astier06}.
The uncertainty in $w$ from gravitational lensing at the $1\sigma$
confidence level for the first year and final SNLS data together with 
BAO constraints is  expected to be at the percent and sub-percent level, 
respectively. Most of this uncertainty, as already noted, should be
included in the intrinsic scatter derived for the SNLS.

\subsection{Removal of outliers}
Since the magnification distribution is asymmetric, it is important to
sample the high magnification tail. Highly magnified SNe~Ia would be
outliers in the Hubble diagram and might therefore be removed. 

\citet{sarkar07} recently quantified the shift in the equation of state
parameter which could be expected due to lensing for future large data sets
($\ga 2000$) of SNe~Ia. They also investigated the effect of removing
bright outliers. According to their simulations, the shift in $w$  
increases from $\sim 0.5\%$ to $\sim 0.8\%$ when bright SNe~Ia deviating
by more than $2.5\sigma$ are removed.

We have performed a simple exercise to investigate the effects of
removing highly magnified SNe~Ia from the Hubble diagram. 
The hatched histograms in Figs.~\ref{fig:ombias} and~\ref{fig:bias} 
show the effect of removing bright $2.5\sigma$ outliers. Approximately
$1\%$ of the SNe~Ia are removed by this cut. The results of the
exercise are given in Table~\ref{tab:bias}.

Let us first discuss the effect of removing bright outliers for case 1.
If bright outliers are removed, the peak of the distribution of 
$\Delta \Omega_{\rm M}$ is shifted to a lower value.
Moreover, the distribution
broadens and become more gaussian.
Since the high magnification SNe~Ia have been removed, most SNe~Ia
appear to be dimmer and consequently the data favour a smaller value
of $\Omega_{\rm M}$. The removal of the asymmetric high magnification
tail also explains why the distribution of $\Delta \Omega_{\rm M}$
becomes more gaussian. The average shift in $\Omega_{\rm M}$ increases
with approximately a factor $3$ when the outliers are removed.

The peak of the distributions shift towards lower values also for case
2. The shift in $w$ increases by a factor approximately $2$ when the
bright outliers are removed. However, the width of the distributions
hardly change at all.

\subsection{Detecting a magnification correlation}
Since amplified and de-amplified SNe~Ia should be brighter and fainter
than average, respectively, we 
expect a correlation between the residuals in the Hubble diagram and the
estimated lensing magnifications, $\Delta_{\mu}$.  
Clearly, intrinsic brightness
scatter among the SNe Ia as well as measurement errors add noise to
the gravitational lensing signal. The ability to detect the
correlation therefore depends on the quality and size of the data set.

\begin{figure}
\resizebox{\hsize}{!}{\includegraphics[angle=-90]{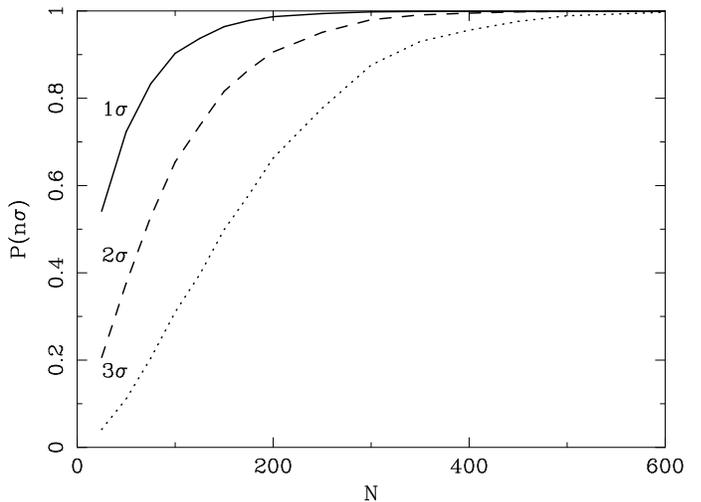}}
\caption{\label{fig:pplot} Probability to detect a correlation between
  estimated magnification and SN~Ia brightness at 
  different confidence
  levels, $P(n\sigma)$, 
  as a function of the number, $N$, of SNe~Ia.
  The solid, dashed, and dotted curves shows the probability to detect a
  correlation at the $1\sigma$, $2\sigma$, and $3\sigma$ confidence
  level, respectively. The simulated data sets have the same
  distributions of redshifts and errors as the first year SNLS data.
}
\end{figure}

Our simulations demonstrate that the quality of the SNLS data is
sufficient for a detection of a correlation at high
confidence. Figure~\ref{fig:pplot} shows the probability of detecting
the correlation at $1\sigma$ ($68.3\%$, solid curve), $2\sigma$
($95.4\%$, dashed curve), and $3\sigma$ ($99.7\%$, dotted curve)
confidence level as a function of the number of SNe~Ia.  For these
simulations we have used the distributions of redshifts and errors from the
first year of SNLS data \citep{astier06}.  If the magnification factor
could be estimated for $450$ SNe~Ia, the probability to detect the
correlation at the 
$3\sigma$ confidence level is
$97.6\%$. 
A firm detection of
the correlation between magnifications and SN~Ia brightnesses is thus
very likely to be found from the final SNLS data. If the magnification
factor could be predicted exactly, the probability 
would instead be
$99.9\%$. 
Even in this best case scenario uncertainties in the Hubble diagram 
residuals would affect our
possibility to find the correlation.

Increasing the intrinsic brightness dispersion from 
$\sigma_{\rm int}=0.13$ mag to $\sigma_{\rm int}=0.15$ mag and
$\sigma_{\rm int}=0.17$ mag would decrease the probability to make a
$3\sigma$ detection of the lensing signal from $97.6\%$ to 
$96.1\%$ and $90.2\%$, respectively. 
The possibility to detect a lensing signal at
high confidence level is thus not critically dependent on the
intrinsic dispersion.

Our chances to make a high confidence level detection of the
correlation using the final SNLS data looks
promising even if our estimates of the lensing masses would be
wrong. The probability to make a $3\sigma$ detection is $\ga 95\%$
even when the lensing masses are over-estimated or under-estimated by 
$\sim 50\%$.
For the final SDSSIISN data set consisting of $\sim 300$ SNe~Ia, on
the other hand, we predict the probability of a $3\sigma$ detection 
to be only $4\%$.

\subsection{Constraints on halo masses}\label{haloconstraints}
\begin{figure}
\resizebox{\hsize}{!}{\includegraphics[angle=-90]{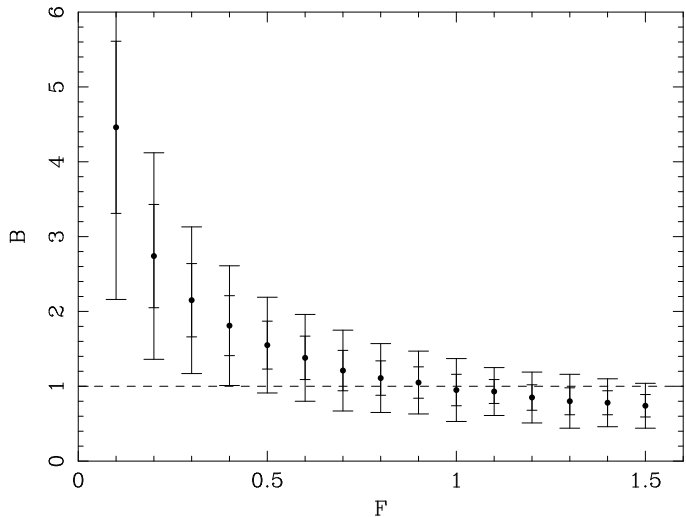}}
\caption{\label{fig:bplot} Slope, $B$, in the magnification-residual 
diagram as a function of the factor $F$ which all halo masses have been
multiplied by. 
The vertical dashed line shows the value expected for a correlation, 
i.e.~$B=1$. The inner and outer errorbars correspond to $1\sigma$ and 
$2\sigma$ confidence level, respectively.
}
\end{figure}

The gravitational lensing signal provided by the correlation between
magnifications and SN~Ia brightnesses should also give some
constraints on masses of the haloes of the lensing galaxies. 
We have performed simulations
where the mass of each halo has been multiplied by a factor $F$. 
The increase or decrease in halo masses is compensated for by 
changes in the
smoothness parameter to ensure that the total mass in the Universe
sums up to the global value.
To
each simulated data set, a linear relation, $\Delta=A+B\Delta_{\mu}$,
was fitted. Figure~\ref{fig:bplot} shows how the slope, described by
the parameter $B$, changes with the value of $F$. The dashed
horizontal line corresponds to $B=1$, which is the value which would
be recovered if the magnification factors were exactly known. If
obvious outliers in the magnification-residual diagram are removed,
the distribution of the $B$ parameter is gaussian.  The inner and outer
error bars in Fig.~\ref{fig:bplot} correspond to $1\sigma$ and
$2\sigma$ errors. The number of SNe~Ia was $N=450$, however the
standard deviation scales roughly as $1/\sqrt{N}$. 

According to Fig.~\ref{fig:bplot} the slope in the
magnification-residual diagram varies with $F$, which should allow us
to probe the normalisation of the halo masses. From the figure we
conclude that the normalisation of the halo masses 
could probably be constrained 
with the final SNLS data set
at the $1\sigma$ and
$2\sigma$ confidence level 
with $\sim 30\%$ and $\sim 60\%$ accuracy, respectively.

\section{Conclusions}\label{discussion}
We have simulated the effect of gravitational lensing in two major
ongoing supernova surveys. For the relatively nearby SDSSII supernova
search, the effect of gravitational lensing will be small. 
For the more distant supernovae in the SNLS survey, we predict that
the signal from gravitational lensing will be observed with high
confidence. Our simulations indicate that a correlation between Hubble
diagram residuals and magnification for individual supernovae will be
present at high (at least $3\sigma$) significance level. This could be
used both to 
somewhat reduce the scatter in the Hubble diagram, 
and to learn about the properties of the lensing material. A project to
investigate this effect in the SNLS data is underway. We also
note that the prospects of using weak lensing of supernovae to
constrain the matter distribution in the Universe with future surveys
such as SNAP look very promising. A satellite like SNAP would provide
not only high redshift SNe~Ia, but also deep observations in many
filters which would allow reliable photometric redshifts of the
galaxies to be obtained.

\begin{acknowledgements}
The Dark Cosmology Centre is funded by the 
Danish National Research Foundation. 
EM and JS acknowledges financial support from the Swedish
Research Council and the Anna-Greta and Holger Crafoord fund.
We thank Reynald Pain for making this project happen. 
We thank Julien Guy for invaluable comments.
We thank Tomas Dahl\'en for helpful discussions concerning photometric 
redshifts and help with conversion of magnitudes.

\end{acknowledgements}


\end{document}